\documentclass[aps,prl,twocolumn,floatfix,superscriptaddress,showpacs,amsfonts
,amssymb,amsmath,preprintnumbers]{revtex4-1}
\usepackage{bm}
\usepackage{epsfig}
\usepackage{amsmath}
\usepackage{amssymb}
\usepackage{graphicx}
\usepackage{bm}
\usepackage{color}
\usepackage{subfigure}

\newcommand{\Fig}[1]{Figure~\ref{#1}}
\newcommand{\fig}[1]{Fig.~\ref{#1}}
 
\newcommand{\eq}[1]{Equation~(\ref{#1})}

\newcommand{\be}{\begin{equation}}
\newcommand{\ee}{\end{equation}}

\newcommand\bea{\begin{eqnarray}}
\newcommand\eea{\end{eqnarray}}

\newcommand{\etal}{\textit{et al.}}

\begin{document}

\title{Magnetic field generation and amplification in an expanding plasma}

\author{K. M. Schoeffler}
\author{N. F. Loureiro}
\affiliation{Instituto de Plasmas e Fus\~ao Nuclear --- 
Laborat\'orio Associado,\\
Instituto Superior T\'ecnico, Universidade de Lisboa, 1049-001 Lisboa, Portugal}

\author{R. A. Fonseca}
\affiliation{Instituto de Plasmas e Fus\~ao Nuclear --- 
Laborat\'orio Associado,\\
Instituto Superior T\'ecnico, Universidade de Lisboa, 1049-001 Lisboa, Portugal}
\affiliation{DCTI/ISCTE Instituto Universit\'ario de Lisboa, 1649-026 Lisboa, Portugal}

\author{L. O. Silva} 
\affiliation{Instituto de Plasmas e Fus\~ao Nuclear --- 
Laborat\'orio Associado,\\
Instituto Superior T\'ecnico, Universidade de Lisboa, 1049-001 Lisboa, Portugal}
\date{\today}

\begin{abstract}
Particle-in-cell simulations are used to investigate 
the formation of magnetic fields, $\mathbf{B}$, in plasmas
with perpendicular electron density and temperature gradients. 
For system sizes, $L$, comparable
to the ion skin depth, $d_i$, it is shown that $B\sim d_i/L$, 
consistent with the Biermann battery effect. However, for large
$L/d_i$, it is found that the Weibel instability (due to electron temperature
anisotropy) supersedes the Biermann battery as the main producer
of $B$. The Weibel-produced fields saturate at a
finite amplitude (plasma $\beta\approx 100$), independent of $L$.
The magnetic energy spectra below 
the electron Larmor radius scale are well fitted by power law with slope $-16/3$, as predicted
in Schekochihin \etal, Astrophys. J. Suppl. Ser. {\bf 182}, 310 (2009).

\end{abstract}


\maketitle


\paragraph{Introduction.}
The origin and amplification of magnetic fields is a central problem in
astrophysics~\cite{Kulsrud08}. The turbulent
dynamo~\cite{Kulsrud92,Brandenburg12} is generally thought to be the basic
process behind the amplification of a magnetic seed field; however, some other
process is required to originate the seed itself. Amongst the few mechanisms
able to do so is the Biermann battery effect, due to perpendicular electron
density and temperature gradients~\cite{Biermann50}. It is often conjectured that
the observed magnetic fields in the universe may be of Biermann origin,
subsequently amplified via dynamo action~\cite{Kulsrud08}.
However, simple theoretical estimates suggest that Biermann-generated magnetic fields, 
$\mathbf{B}$, should be such that~\cite{Max78,Craxton78,Haines97}
\be
\label{Biermann_scaling}
\beta\equiv 8\pi P/B^2\sim \left(d_i/L\right)^{-2},
\ee 
where $P$ is the plasma pressure, $d_i=c/\omega_{pi}$ is the ion inertial length 
(with $c$ the speed of light and $\omega_{pi}$ the ion plasma frequency)
and $L$ is the characteristic length scale of the system. Given the
extremely small values of $d_i/L$ typical of astrophysical systems, it is
an open question whether such seeds are sufficiently large to account for the
microgauss fields observed today.

Megagauss magnetic fields are observed to form in intense laser-solid
interaction laboratory experiments~\cite{Stamper71,Nilson06,Li06,
Li07,Kugland12}. In these experiments, the laser generates an expanding bubble
of plasma by ionizing a foil of metal or plastic. The plasma is denser closer
to the plane of the target foil, and hotter closer to the laser beam axis.
Perpendicular density and temperature gradients are thus generated, giving rise
to magnetic fields via the Biermann effect. Besides their intrinsic interest,
these experiments offer a unique opportunity to illuminate a fascinating, and
poorly understood, astrophysical process.

In this Letter we perform {\it ab initio} numerical investigations of the
generation and growth of magnetic fields in a configuration akin to that of
laser-generated plasma systems. For small to moderate values of the parameter
$L/d_i$ our simulations confirm the theoretical predictions of
Haines~\cite{Haines97}; in particular, for $L/d_i\gtrsim1$ the magnetic fields
obey the scaling of \eq{Biermann_scaling}. However, when $L/d_i\gg 1$, we find
that the plasma is unstable to the Weibel instability~\cite{Weibel59}, which
amplifies the magnetic fields such that $\beta\approx 100$, independent of $L$.
These results have strong implications for the interpretation of laser-solid
interaction experiments; they also shed new light on the currently accepted
view of the origin of the observed cosmic magnetic fields. 

\paragraph{Computational Model.}
We perform a set of particle-in-cell (PIC) simulations using the OSIRIS
framework~\cite{Fonseca02,Fonseca08}.  The initial fluid velocity, electric
field, and magnetic field are all uniformly zero. We start with a spheroid
distribution of density, that has a shorter length scale in one direction: 
    $n = (n_0-n_b) \cos(\pi R_1/2L_T) + n_b, \mbox{ if } R_1 < L_T, n_b, \mbox{ otherwise},$
where $R_1 = \sqrt{x^2+(L_T/L_n y)^2+z^2}$
and $n_b=0.1n_0$ is the uniform background density. The characteristic lengths
of the temperature and density gradients generated by the laser beam are
denoted by $L_T$ and $L_n$, respectively. To represent the recently ionized
foil, which is flatter in the direction of the laser, $y$, we set $L_T/L_n =
2$. (This is a generic choice that appears to be qualitatively consistent with
experiments, e.g.~\cite{Nilson06,Li06,Li07,Kugland12}; 
note, however, that the specific value of $L_T/L_n$
depends on target and laser properties.)
ion thermal velocity $v_{Ti0}$. The spatial profile for the electron thermal
velocity is cylindrically symmetric along the $y$ direction, where it is
hottest in the center:
$v_{Te} = (v_{Te0}-v_{Teb}) \cos(\pi R_2/2L_T) + v_{Teb}, \mbox{ if } R_2 < L_T, v_{Teb}, \mbox{ otherwise},$
where $R_2 = \sqrt{x^2+z^2}$,
resulting in a maximum initial electron pressure $P_{e0} = m_e n_0
v_{Te0}^2/2$. The numerical values of the thermal velocities are
$v_{Te0}=0.2c$ and $v_{Ti0}=v_{Teb}=0.01c$.
Note that in our setup the pressure is
dominated by the electrons, and thus $\beta\approx\beta_e \equiv 8\pi P_e/B^2$.
For simplicity the boundaries are periodic, but the box is large enough that
they do not interfere with the dynamics
[$L_{\left(x,y,z\right)max}=-L_{\left(x,y,z\right)min}= 15/8 L_T$]. In order to
investigate a larger range of $L_T/d_i$, the simulations are run with a reduced
mass ratio of 25. The spatial resolution is $16$ gridpoints/$d_e$, or $2.26$
gridpoints/$\lambda_d$, where $d_e=c/\omega_{pe}$ is the electron inertial 
length ($\omega_{pe}$ is the electron plasma frequency) and $\lambda_d$ is the 
electron Debye length. 
The time resolution is $\Delta t\omega_{pe}= 0.07$.
The 2D simulations have
196 or 64 particles per cell (ppc); the 3D simulation has 27 ppc.

\paragraph{Biermann regime.}

\begin{figure}
  \noindent\includegraphics[width=4.0in]{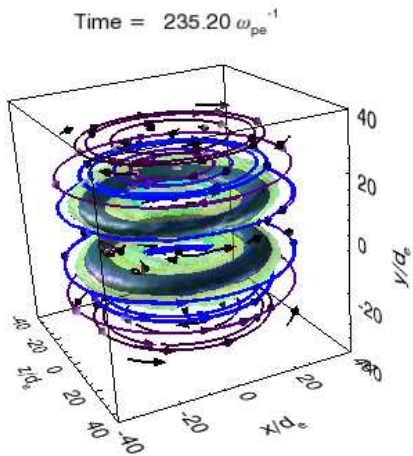}
  \caption{\label{biermann3d}
    Magnetic energy contours after saturation ($t\omega_{pe} = 235.2$, see \fig{timetraces}) 
    from a 3D simulation with $L_T/d_e=50$. Lighter to darker colors represent
    $B^2/8\pi P_{e0} = 0.0035,0.0071,0.0106$.
    Several magnetic field lines are also displayed.
  }
\end{figure}

\begin{figure*}
  \centering
  \includegraphics[width=0.4\textwidth]{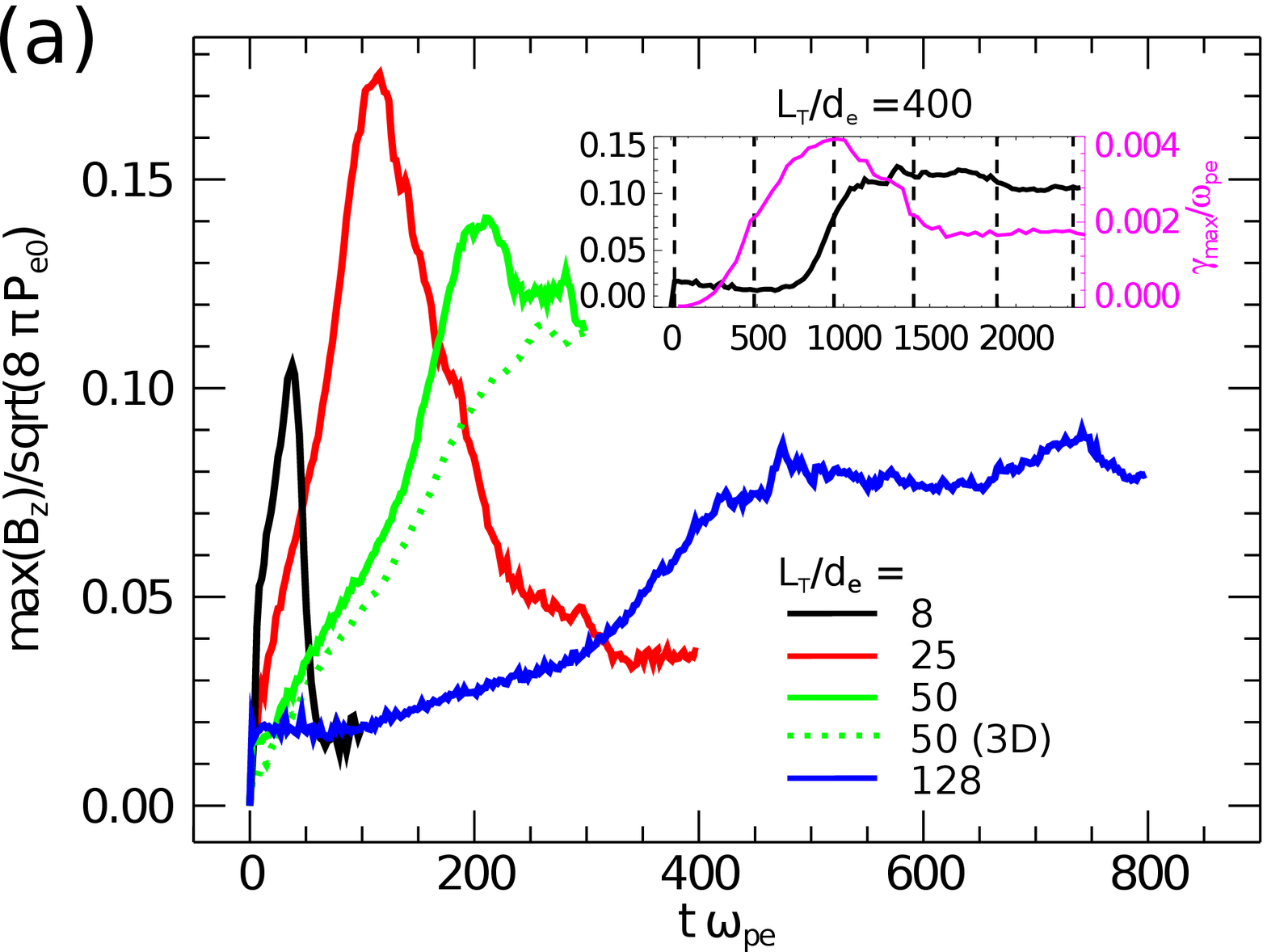}
  \hskip1cm
  \includegraphics[width=0.4\textwidth]{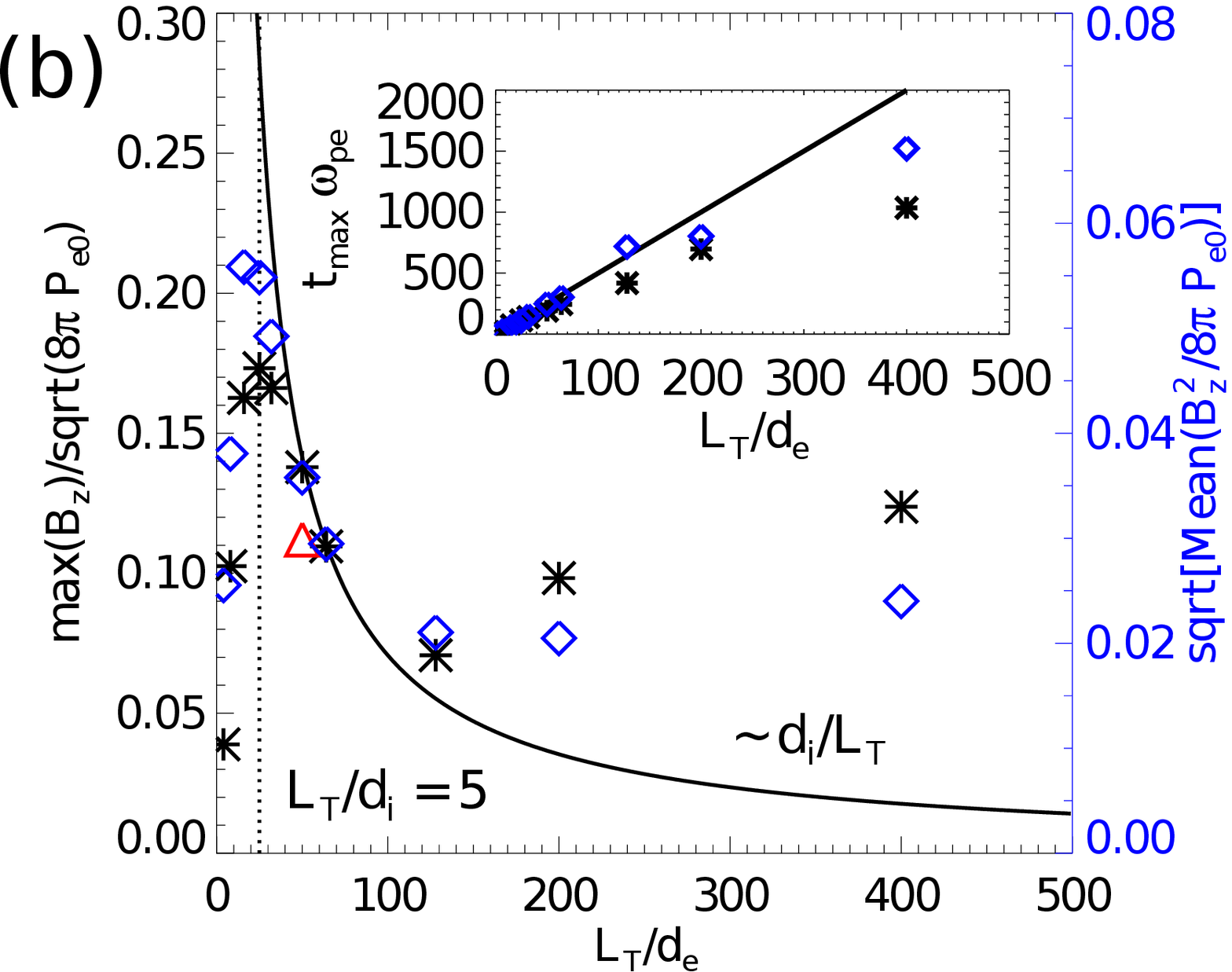}
  \caption{\label{timetraces}
    (a) Maximum $B_z$ {\it vs.} time for a selection of system sizes
    ($L_T/d_e$).  The inset shows the $L_T/d_e=400$ case (black). The magenta line
    is the maximum Weibel growth rate, $\gamma_{max}$, at $y=0$. Dashed lines
    identify the times at which the spectra of \fig{kspect} are calculated. 
    (b) Maximum (asterisks), and average magnitude (diamonds) of
    $B_z$ {\it vs.} $L_T/d_e$. The triangle represents the
    maximum $B_z$ for the 3D run. The solid curve is $
    \max(B_z)/\sqrt{8 \pi P_{e0}} = \sqrt{2}d_i/L_T$; the
    dotted line indicates $L_T/d_i = 5$. The inset shows the 
    time to maximum magnetic field, $t_{max}$,
    {\it vs.} $L_T/d_e$. The solid line indicates $t_{max}=L_T/v_{Te0}$.
  }
\end{figure*}

\Fig{biermann3d} shows contours of constant magnetic energy density and
magnetic field lines from a 3D simulation with $L_T/d_e=50$ taken at
$t\omega_{pe}=235.2$, after the magnetic field strength saturates (see
\fig{timetraces}). As expected based on the initial conditions, we observe the
formation of large-scale azimuthal Biermann magnetic fields which are nearly
axisymmetric. Although Biermann generation of magnetic fields has been
investigated before~\cite{Thomas12}, this is the first fully self-consistent
kinetic 3D simulation.

\begin{figure}
  \noindent\includegraphics[width=3.0in]{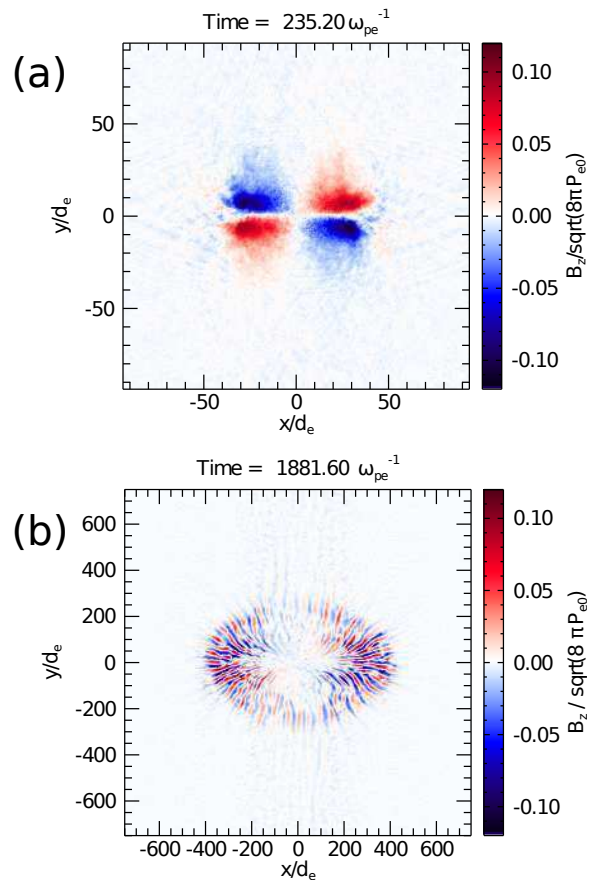}
  \caption{\label{combined2d}
    Out-of-plane magnetic field, $B_z$, after saturation (see \fig{timetraces}) 
    for (a) $L_T/d_e = 50$,
    and (b) $L_T/d_e = 400$.
  }
\end{figure}

The axisymmetry in the 3D simulation suggests that a scaling study in system
size can be performed using a more computationally efficient 2D setup. To this
end, we take a cut of the 3D system at $z=0$, where the azimuthal
(out-of-plane) magnetic fields are in the $z$ direction, and perform a set of
2D simulations with $L_T/d_e= (4,8,16,25,32,50,64,128,200,400)$. For $4\le
L_T/d_e \le 128$ we use 196 ppc. For $L_T/d_e=200,400$ we use 64 ppc instead
due to computing time limitations; convergence studies at lower values of
$L_T/d_e$ do not show significant differences between 196 and 64 ppc. A
snapshot taken at the same time of a 2D version of the simulation presented in
\fig{biermann3d} is shown in \fig{combined2d}(a) for comparison. The same
large-scale magnetic field structure is manifest, with very similar levels of
$B_z$.

The time trace of the maximum magnetic field strength for a selection of cases
can be seen in \fig{timetraces}(a). For small systems, $L_T/d_e < 50$, the
magnetic field reaches a maximum and then decays away. On the other hand, we
observe that for $L_T/d_e > 50$ the magnetic field saturates at around its peak
value.

\begin{figure}
  \noindent\includegraphics[width=3.0in]{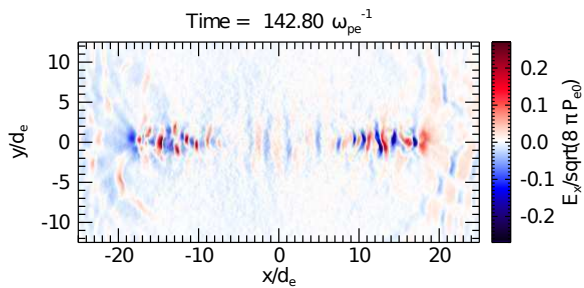}
  \caption{\label{IAI}
    Electric field in the $x$-direction, $E_x$, for
    $L_T/d_e=25$ at $t\omega_{pe}=142.8$.
  }
\end{figure}

\Fig{timetraces}(b) shows the scaling with system size of the maximum and the
average magnitude of the magnetic field (the square root of $B_z^2$ averaged in
a box $2L_T\times2L_n$ surrounding the expanding bubble) at the time when the
field saturates (or peaks for $L_T/d_e<50$).  There are three distinct regions
in this plot. For $L_T/d_e<25$ (i.e., $L_T/d_i\lesssim 5$), the magnetic field
increases with system size. This stage is followed by a region where the
saturated amplitude of the field decreases as $d_i/L_T$, which lasts while
$L_T/d_e < 100$. These two stages confirm the theoretical prediction of
Haines~\cite{Haines97}: in very small systems, there is a competition between
the Biermann battery effect and microinstabilities (the ion acoustic and the
lower hybrid drift instabilities), triggered by an electron drift velocity in
excess of the ion acoustic speed, which suppress the Biermann fields. As the
system becomes larger, the electron drift velocity decreases. (Larger systems
have larger-scale magnetic fields, and therefore lower currents.) The
microinstabilities thus become progressively less virulent until their complete
suppression, whereupon we encounter a ``pure'' Biermann regime, as described in
\eq{Biermann_scaling}. Inspection of the simulations for $L_T/d_e<50$ at times
after the magnetic field reaches its peak value shows clear electric field
perturbations along $y=0$, consistent with the ion acoustic instability.  These
are exemplified for $L_T/d_e=25$ in \fig{IAI}. Note that the density gradient
goes to zero at $y=0$, ruling out the lower hybrid drift instability as the
cause of the decay of the magnetic field.

\paragraph{Weibel regime.}
An unexpected third regime is encountered for $L_T/d_e>100$. In that region of
\fig{timetraces}(b), the magnetic field produced in our simulations no longer
follows the predicted $d_i/L_T$ Biermann scaling, but rather increases with the
system size and appears to tend to a constant, finite value, $\beta_e \approx
100$.

In this new regime the magnetic fields are produced by the Weibel
instability~\cite{Weibel59}. The initial cloud of plasma expands due to the
imposed density gradient, generating both outward ion and hot electron flows.
The velocity of the electron flows vary along the temperature gradient. The
higher temperature flows originating in the center, stream past lower
temperature inward flows originating further outward, which maintain
quasineutrality. This generates a larger velocity spread (larger temperature)
in the direction of the flow, while the perpendicular temperature remains
unaffected. It is this temperature anisotropy that drives the Weibel
instability~\cite{Weibel59}. Note that along $x=0$, where the temperature
gradient is zero no anisotropy is generated, and thus the Weibel instability is
not observed [see \fig{combined2d}(b)].

As exemplified in \fig{combined2d}(b) for our largest simulation ($L_T/d_e =
400$), the large-scale coherent Biermann magnetic fields characteristic of the
smaller systems are replaced by non-propagating magnetic structures with very
large wavenumbers ($k d_e \sim 0.2$), and with a transverse wave vector,
$\mathbf{k}$, perpendicular to the direction with a larger temperature. These
features are consistent with the Weibel
instability~\cite{Weibel59,Califano98,Fonseca03}. In addition we have
compared our results with the analytic growth rate predicted by
Weibel~\cite{Weibel59}. In our simulations we observe an enhanced temperature
in the direction of the density gradient (parallel) as high as $A \equiv
T_{\parallel e}/T_{\perp e} -1 \approx 2.0$. In a cut at $y=0$ we calculate the
Weibel growth rate, $\gamma$, for the fastest growing $k$, $k_{max}$, using the
locally measured values of $n$, $T_{\perp e}$, and $A$. The maximum $\gamma$ of
this cut, $\gamma_{max}$, is plotted vs. time in \fig{timetraces}(a), showing a
peak when the magnetic field strength rises exponentially, and a subsequent drop
corresponding to the loss of anisotropy after saturation. The magnitude of the
growth rate thus calculated is also consistent within a factor of 2, with
$k_{max}d_e\approx0.2$, analogous to the structures in \fig{combined2d}(b).


The transition between the Biermann and Weibel regimes is also visible in the
inset plot in \fig{timetraces}(b), where we show the time to reach the maximum
magnetic field, $t_{max}$, as a function of system size. For $L_T/d_e < 50$, we
find that $t_{max}\sim L_T/v_{Te0}$. A linear in time scaling is indeed to be
expected for Biermann generated fields; also, at these small scales the
electrons are not coupled to the ions, and are thus free to move at their
thermal velocity. A transition to a logarithmic dependence on the system size
occurs after $L_T/d_e > 50$; this is expected since the Weibel instability
amplifies the magnetic fields at an exponential rate. Note that the Weibel
instability cannot occur below a certain system size because it is suppressed
by the strong, large-scale Biermann fields. (We have confirmed this suppression
numerically by running a similar setup where the Biermann effect is not
present; see also~\cite{Molvig75}.) 

\begin{figure}
  \noindent\includegraphics[width=3.0in]{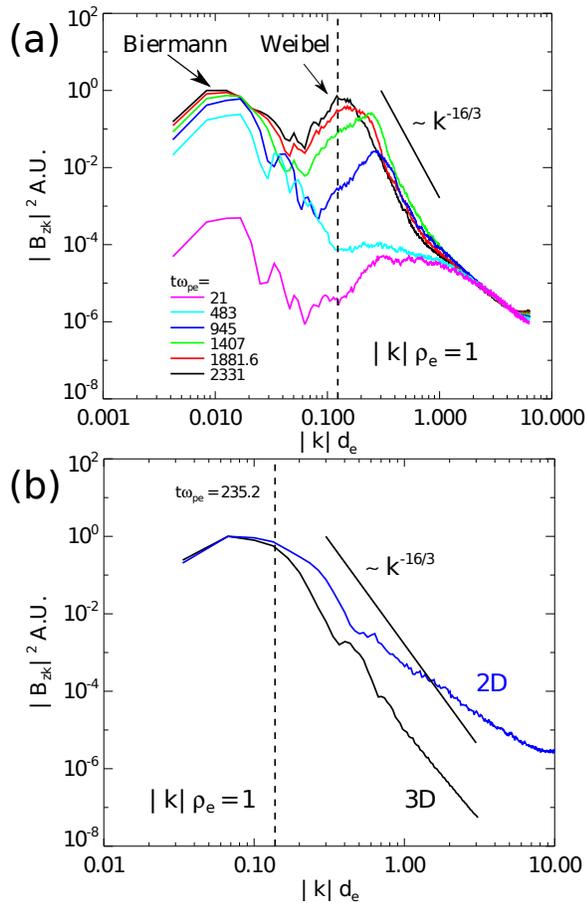}
  \caption{\label{kspect} Fourier spectrum of 
    $B_z^2$ for (a) $L_T/d_e=400$, and (b) $L_T/d_e=50$. 
    In (a) the spectrum is shown at several different times [see \fig{timetraces}(a)]
    while in (b) the 3D (black
    curve) and the 2D simulation (blue curve) are shown for $t\omega_{pe}=235.2$.
    The dashed lines represent where $k\rho_e =1$, based on the maximum magnetic
    field. The solid black lines indicate a power law of $k^{-16/3}$.
  }
\end{figure}

We have performed additional studies that confirm our conclusions up to a
mass-ratio of $m_i/m_e = 2000$, at which point the results have converged. With
these more realistic mass ratios, the saturated magnetic field increases less
than twice the value obtained for $m_i/m_e=25$. These results will be presented
elsewhere.

\paragraph{Spectra.}
\fig{kspect}(a) shows the spectrum of $B_z^2$ for our largest simulation
($L_T/d_e=400$) at the times indicated in the inset of \fig{timetraces}(a). At
early times a peak rapidly forms at $kd_e \approx 0.01$, which corresponds to
the large-scale Biermann-generated magnetic field. At later times, a second
peak corresponding to the Weibel generated magnetic fields begins to form at
$kd_e \approx 0.2$ and eventually saturates at $kd_e\approx0.1$; this scale
corresponds to $k\rho_e=1$, where $\rho_e$ is the electron Larmor radius based
on the maximum $B_z$ at saturation. Therefore, the Weibel generated fields
saturate when $\beta_e=(\rho_e/d_e)^2\approx 100$
(cf.~\cite{Califano98},~\cite{Romanov04}), independent of the system size as
shown in \fig{timetraces}(b). 

Another remarkable feature yielded by the spectra of \fig{kspect} is the power
law behavior of the magnetic energy at sub-$\rho_e$ scales, with a slope close
to $-16/3$. A less steep power law appears to exist at smaller scales, but this
is not present in the 3D simulation, as seen in \fig{kspect}(b). Note that this
slope occurs for both small and large systems and is not, therefore, a
consequence of the Weibel instability. Such a power law dependence was
theoretically predicted using gyrokinetic theory in~\cite{Schekochihin09},
where it was identified as resulting from an entropy cascade of the electron
distribution function at scales below $k\rho_e\sim 1$.  We believe this is the
first 3D confirmation of that prediction, although similar observations have
been made in 2D simulations~\cite{Camporeale11}.

\paragraph{Conclusions.}
We have performed fully kinetic simulations of magnetic field generation and
amplification in expanding, collisionless, plasmas with perpendicular density
and temperature gradients. For relatively small systems, $L_T/d_e < 100$, we
observe the production of large-scale magnetic fields via the Biermann battery
effect, fully confirming the theoretical predictions of Haines~\cite{Haines97},
in particular the scaling of the magnetic field strength with $d_i/L_T$. For
larger systems, however, we discover a new regime of magnetic field generation:
the expanding plasmas are Weibel unstable, giving rise to small scale ($k d_e
\sim 0.2$) magnetic fields whose saturated amplitude is such that
$\beta_e\approx 100$, independent of system size, and thus much larger than
would be predicted for such systems on the basis of the Biermann mechanism.  We
note that both of these regimes can in principle be probed by existing
experiments. For example, the $L_T/d_i\approx 1$ regime (Biermann) is
accessible to the Vulcan laser~\cite{Nilson06}, whereas $L_T/d_i\approx100$
(Weibel) is reachable by an OMEGA laser~\cite{Li06}. In practice,
however, collision frequencies large compared to the electron transit time
prohibit electron temperature anisotropies, thereby inhibiting the Weibel
instability.  If less collisional regimes can be attained in the experiments,
it may be possible to experimentally investigate the transition from Biermann
to Weibel produced magnetic fields that we have uncovered here.  

In the context of (largely collisionless) astrophysical plasmas, our results
may significantly impact the canonical picture of cosmic magnetic field
generation~\cite{Kulsrud08}, by suggesting that Biermann seed fields may be
pre-amplified exponentially fast via the Weibel instability up to reasonably
large values (i.e., independent of the system size) previous to turbulent
dynamo action. 

\paragraph{Acknowledgments.}
This work was partially supported by Funda\c{c}\~ao para a Ci\^{e}ncia e
Tecnologia (Ci\^encia 2008 and Grant nos. PTDC/FIS/118187/2010 and
PTDC/FIS/112392/2009), by the European Research Council (ERC-2010-AdG Grant no.
267841), and by the European Communities under the Contract of Association
between EURATOM and IST. Simulations were carried out at Kraken, NICS (XSEDE
Grant AST030030), and at SuperMUC (Germany) under a PRACE award.


%

\end{document}